\documentclass[manuscript]{acmart}

\usepackage{xcolor}
\usepackage{lscape}

\begin{document}

%

\title{Computing-specific pedagogies and theoretical models: common uses and relationships}

\author{Lauri Malmi}
  \orcid{0000-0003-1064-796X}
  \affiliation{
    \institution{Aalto University}
    \country{Finland}}
  \email{lauri.malmi@aalto.fi}
\author{Judy Sheard}
  \orcid{0000-0002-4179-8149}
  \affiliation{
    \institution{Monash University}
    \country{Australia}}
  \email{judy.sheard@monash.edu}
\author{Claudia Szabo}
  \orcid{0000-0003-2501-1155}
  \affiliation{
    \institution{The University of Adelaide}
    \country{Australia}}
  \email{claudia.szabo@adelaide.edu.au}
\author{P\"{a}ivi Kinnunen}
  \orcid{0000-0002-8650-4925}
  \affiliation{
    \institution{University of Helsinki}
    \country{Finland}}
  \email{paivi.kinnunen@helsinki.fi}

\renewcommand{\shortauthors}{Malmi, Sheard, Szabo, Kinnunen}

\begin{abstract}
Computing education widely applies general learning theories and pedagogical practices. However, computing also includes specific disciplinary knowledge and skills, e.g., programming and software development methods, for which there has been a long history of development and application of  specific pedagogical practices. In recent years, there has also been substantial interest in developing computing-specific theoretical models, which seek to describe and explain the complex interactions within teaching and learning computing in various contexts. In this paper, we explore connections between computing-specific pedagogies and theoretical models as reported in the literature. Our goal  is to enrich computing education research and practice by illustrating how explicit use of field-specific theories and pedagogies can further the whole field.        We have collected a list of computing-specific pedagogical practices and theoretical models from a literature search, identifying source papers where they have been first introduced or well described. We then searched for papers in the ACM digital library that cite source papers from each list, and analyzed the type of interaction between the model and pedagogy in each paper. We developed a categorization of how theoretical models and pedagogies have supported or discounted each other, have been used together in empirical studies or used to build new artefacts. Our results showed that pair programming and parsons problems have had the most interactions with theoretical models in the explored papers, and we present findings of the analysis of these interactions.  

\end{abstract}

%
%



%
\maketitle

%

\section{Introduction}
\label{sec:intro}

Computing Education Research (CER) is a relatively young, evolving field of science. Recent research has highlighted the ways the field has developed 
\cite{simon2015emergence,tedre2018changing,apiola2022computing},
 with further studies focusing on new computing education-specific theories ~\cite{malmi2019CompEdTheories,malmi2020TheoriesOfEmotion,malmi2022Developing}.
 The emergence of field-specific theories and models is regarded as a sign of a maturing research field~\cite[Chapter 1]{Fensham2004DefiningIdentity}.  This paper positions itself into this same line of research by investigating the connections between the emerging computing-specific \emph{theoretical aspects} of learning and \emph{pedagogical practices} in computing classrooms and other educational contexts. Educational theory seeks to better understand teaching and learning related phenomena to find ways to support students’ learning and we argue that such theories should inform development of practice.  The amount and depth of inter-connections between computing-specific theories and pedagogies thus highlight the evolution of CER as a research field, specifically whether it has matured sufficiently to produce theories and pedagogical practices that impact the development of each other.



When considering the quality of teaching and learning, the importance of the relationship between learning theory and pedagogical practice is evident.  The knowledge of general and field-specific learning theories are a focal part of a teacher's pedagogical content knowledge, which is also grounded in knowledge about how a particular topic could be taught \cite{shulman1987knowledge}. Learning theories provide teachers with some basic tenets on which to  build their pedagogical choices. For instance, pedagogies, such as problem-based learning that require students to work together and question what they already know about the topic and what still needs to be learned in order to complete the task at hand, are well in line with the constructivist learning theory. 
General learning theories are widely applicable in computing education, and there is a large corpus of research addressing pedagogical practices in computing education that is based on general educational research literature, e.g.,~\cite{margulieux2019learning,robins2019cognitive,falkner2019Pedagogy,lishinski2019motivation}.   However, the computing discipline also has its own specialized disciplinary knowledge and skills, e.g., programming, which are unique and may be best taught with computing-specific pedagogical practices, informed also by computing-specific theoretical models. 
The complete mechanisms of pedagogical influence are currently far from being fully understood, and much more theoretical and empirical work is needed; indeed, educational settings are so diverse and complex that building accurate predictive theories is a rare option, if possible at all.
\citet{tedre2022grand} therefore argue for the development of models in educational settings~\footnote{They discuss in depth the complexity of interpreting the concept "theory".  Such discussion is, however, beyond the scope of this paper, and we therefore use terms "theory", "model" or "theoretical model" without denoting any specific difference between them.}. Theoretical models, which identify relevant factors and their interaction can, however, be very useful to better understand the complexities of teaching and learning processes and support systematic development of teaching. In this respect, the results of this study aim to contribute to our collective pedagogical content knowledge by bringing forward the interplay of emerging field-specific theories and pedagogies of learning computing. We are not aware of any survey of such theory-pedagogy relationships focusing on computing education.

Several large scale reviews have identified substantial research on building theoretical models of teaching and learning computing~\cite{lishinski2016methodologicalRigor,malmi2019CompEdTheories,malmi2020TheoriesOfEmotion,malmi2022Developing}. Malmi et al. also explored whether computing-specific theories had informed teaching practice in computing education finding some limited evidence of their impact~\cite{malmi2023domain_practice}. However, their point of view focused on whether a theory was "Used to design a new pedagogical method" [Ibid, p.5].
In this paper, we seek to look at the theory-pedagogy interaction from a wider perspective. Theories can also provide support for teachers' pedagogical choices by giving insight into how the pedagogy influences the learning process and theories can systematically further the development of pedagogies by giving arguments as to why an existing pedagogy is successful or not. On the other hand, a pedagogy can support a theory by giving evidence of whether the specific implementation of the learning process matches  with the theoretical explanation. Moreover, both theory and pedagogy can be used to discuss and analyze findings in a complex situation. Thus, we set our goal to explore this interaction much more broadly than has been carried out in earlier research~\cite{malmi2023domain_practice}.  Our focal interest in computing-specific theories and pedagogies stems from our view that the field -- computing education research (CER) -- should develop its own deep understanding of how teaching and learning of computing takes place, what factors are involved and how practice is informed of their role.

\section{Computing-education-specific theories and pedagogies}
\label{sec:related}

Each discipline has its own characteristics and challenges for learning, and computing, with its roots in three different research traditions~\cite{tedre2008three}, is no different.  First, the \textit{mathematics} tradition emphasizes formal presentation, coherent theoretical structures, and creation and proof of hypotheses and theorems. Much research in theoretical computer science, algorithms, and machine learning follows this tradition. Second, a significant part of computing follows the \textit{scientific} tradition, which builds on forming hypotheses, constructing models and making predictions based on these, designing experiments, and collecting data and analyzing results, in an iterative approach until the model is sufficiently accurate.~\cite{denning1989computing}.   Much of computing education research builds on this tradition.
The third tradition in computing follows an \textit{engineering} approach. The construction of software and hardware artifacts has always been a core computing activity, with design as its foundation block, as \citet[p.64]{denning1989computing} formulated: "Design is the bedrock of engineering: engineers share the notion that progress is achieved primarily by posing problems and systematically following the design process to construct systems that solve them."  

This richness of the computing tradition poses significant challenges for computing educators as 
very different pedagogical methods are needed in various courses, e.g., formal mathematical analysis, theorem development and proving, as opposed to software development in large-scale projects. 
Furthermore, many, perhaps most, core computing concepts are abstract in the sense that there is no obvious corresponding  "real world" physical object that would be familiar to students in their everyday life.  Teachers can build some of these connections between "abstract" and "real", as is carried out, for example, in CS unplugged pedagogy~\cite{battal2021computer} and tangible computing~\cite{horn2019tangible}. Finally, learning computing is not just about learning concepts and their relationships, but also involves learning many abstract skills, such as conceptual analysis, problem solving, programming, testing, debugging, software design.  It is thus apparent that computing-specific pedagogical methods are needed to complement generic pedagogical methods.

\citet{shulman2005signature} presents the concept of \emph{signature pedagogy}, which denotes pedagogical practices that support the preparation of students for a specific profession, for example, bedside teaching at medical schools or case dialogues at law schools. In a signature pedagogy the goal is to learn about ways of thinking within the profession beyond working practices. Shulman distinguishes three levels of a signature pedagogy. First, there is \emph{surface structure}, i.e., concrete, operational acts of teaching and learning, such as showing, demonstrating, questioning, interacting, etc.  Second, there is \emph{deep structure}: "assumptions about how best to impart a certain body of knowledge and know-how", and finally there is \emph{implicit structure}, "a moral dimension that comprises a set of beliefs about professional attitudes, values, and dispositions."\cite[p55]{shulman2005signature}.  

When considering computing education, an example of a signature pedagogy could be pair programming, where students familiarize themselves with programmers' working methods: working in pairs and alternating between the driver and navigator roles (surface structure) during a coding session. While doing this, they learn to read, write, discuss and critique code (deep structure) and learn about professional values: code quality matters (implicit structure).  On the other hand, not all computing-specific pedagogical practices are signature pedagogies. for example, Parsons problems~\cite{Parsons2006parson}, is a technique which supports learning to code by simplifying the coding task into building a puzzle of program statements instead of writing all the code. While this method has demonstrated positive impact on learning~\cite{ericson2022Parsons}, it is not a professional practice.

\citet{falkner2019Pedagogy} present a comprehensive overview of pedagogical approaches in computing education.  They identify and discuss six \emph{pedagogical approaches}, i.e., broader guidelines to implement teaching, each of which may cover a number of specific \emph{pedagogical practices}, denoting a specific activity within a course.  These are \emph{Active learning, Collaborative learning, Cooperative learning, Contributing student pedagogy}, and \textit{Blended learning and MOOC}. Many of the example practices are generic, as one would expect, such as flipped classroom as an activity implementing blended learning, or content creation as an activity implementing a contributing student pedagogy.  However, they also present several examples of computing-specific activities, such as, live coding or test-driven development as instances of active learning or pair programming as an instance of cooperative learning.  

\citet{sanders2017folk} explored the concept of \emph{active learning} by surveying the computing education community as well as carrying out a systematic literature review to find out how active learning is interpreted and what kind of practices are associated with it. Their results implied that while active learning is widely considered a good thing, it is often not a well defined approach, and it is rarely connected to learning theories.  They identified 38 different activities that implement active learning in some way, and classified them into five broader categories: Lecture content outside the lecture, Activities during lecture time, Collaboration and social engagement, Techniques from other disciplines, and Software development techniques. They also identified a number of techniques focused on the change in the instructor's perspective of teaching rather than specific student activities. 

The use of theories in computing education research has received substantial attention in recent years, with several reviews and special issues focusing on this theme. \citet{Malmi2014TheoreticalUnderpinnings} analyzed papers published in the ICER conference, and two major journals, ACM Transactions on Computing Education (TOCE) and Computer Science Education (CSE) from 2005-2011, identifying the use of a large number of theories, models and frameworks; although, these constructs were only found in only roughly half of the papers.  They also analyzed the origin discipline of the constructs, where the largest groups had their origin in educational sciences, psychology and computing. A small set of constructs had been developed in computing education research. \citet{lishinski2016methodologicalRigor} complemented this analysis by looking at papers published in ICER and CSE in 2012-2015 identifying significantly higher number of papers using theory outside CER than in the earlier review~\cite{Malmi2014TheoreticalUnderpinnings}. A few years later, \citet{szabo2019review} carried out an extensive review of the use learning theories in the CER literature.  They collected a large set of general learning theories from~\cite{learningtheories} augmenting it with theories from~\cite{Malmi2014TheoreticalUnderpinnings,malmi2019CompEdTheories} and searched ACM Digital Library for their occurrences in papers published in different venues.  They closely analyzed the key actors in a set of theories, how they interacted and what internal aspects (cognitive, affective, behavioral) were involved. A further analysis investigated how theories were connected with each other by determining where two theories occurred in the same publication and using this to identify three clusters of theories that worked together: experiental theories, theories of mind. and social theories. This work was continued by \citet{szabo2023LearningTheories}, who investigated more closely the interaction of theories. Looking at papers that cited two theories, they presented a more detailed clustering of learning theory communities:  behaviourist and cognitivist learning theories, working memory and experiential learning theories, motivation and behaviourism learning theories, as well as a computing education learning theories community. They investigated the computing education learning theories community more closely, by categorizing the type of interactions the pairs of theories had in the papers. The categories ranged from cases with just casual mentions to cases where theories were discussed separately or together or they were critically compared, and further to cases when both theories were used to analyze results or inform some design or finally the build or extend some artefact or theory.

Computing-specific learning theories or models have also been explored Malmi et al. (calling them domain-specific theories) in multiple papers~\cite{Malmi2014TheoreticalUnderpinnings,malmi2019CompEdTheories,malmi2020TheoriesOfEmotion,malmi2022Developing}. They identified in total 124 theoretical constructs (including theories, models, frameworks and theory-based instruments) in eleven different focus areas in computing education research: assessment/self-assessment, computing education research, content/curriculum/learning goals, emotions/attitudes/beliefs/self-efficacy, errors/misconceptions, learning/understanding, learning behavior/strategies, perceptions of computer science/computing, performance/progression/ retention, study choice/orientation and teaching/pedagogical content knowledge~\footnote{See~\cite{malmi2022Developing} for their definition of these areas.}. They analyzed the methods used to develop the constructs, for what purposes they had been developed, and how they had been used in further research, and citing the original papers where the constructs had been presented. 
In their later work, they analyzed whether these computing-specific theoretical constructs had had any impact on developing or suggesting new pedagogical practices, and found very little evidence of this.~\cite{malmi2023domain_practice}.


\section{Research Questions and Method}
\label{sec:method}

In this work, we seek to explore more widely and deeply the connections between computing-specific pedagogies and theoretical developments.  \citet{malmi2023domain_practice} explored whether computing-specific theories and models had informed new pedagogical developments, which were presented either in papers that cite the theory paper or as pedagogical implications in the discussion part of the theory paper itself.  However, theories and pedagogies can interact in many different ways, e.g., as shown by \citet{szabo2023LearningTheories} between different computing-specific theories. 

To better understand relationships between computing-specific pedagogies and theoretical developments, and given the lack, to the best of our knowledge, of any broad analysis of computing-specific pedagogies, our first goal is  to identify a set of such pedagogies, as reported in the CER literature, followed by an identification of potential locations of theory-pedagogy relationships and an analysis of relationship depth. Our research questions are:

\begin{enumerate}
    \item RQ1: What computing-specific pedagogies are defined in the CER literature? 
    \item RQ2: What computing-specific learning theories and models~\footnote{As mentioned earlier, in this paper the distinction between the nature of theories and models is not relevant to the discussion. We use the term theory for denoting both of them.}
    identified by a name are discussed in the CER literature?
    \item RQ3: What kind of connections have been made between computing-specific theories/models and pedagogies in CER?
\end{enumerate}

To better understand the results to the third research question, we focused on the cases with many connections between theories and a specific pedagogy. The two pedagogies with the most connections were pair programming and Parsons problems, and we therefore added two subquestions.

\begin{enumerate}
    \item RQ3.1: What type of connections have been made between computing-specific theories and pair programming? 
    \item RQ3.2: What type of connections have been made between computing-specific theories and Parsons problems? 
\end{enumerate}



Data was collected and analyzed in several phases, as discussed below. 

\subsection{Identifying computing-specific pedagogies}

Pedagogy is a concept that has no clearly agreed definition. In general, it can be interpreted as a methodological approach, technique or activity that teachers apply to support their students to achieve certain learning goals or other objectives, such as increasing motivation or engagement.  A pedagogy can cover a holistic process, which integrates several techniques over a longer period.  For example, in project-based learning, students can work in teams to address a complex open problem during a whole semester using several practices, such as team discussions, self-studying, following supplementary lectures, and designing and/or building some artefacts and writing reports. On the other hand, a pedagogy can cover specific techniques applied in a classroom for a limited time, such as analyzing fading worked examples given by a teacher for a specific topic.

Numerous pedagogical methods are generic in the sense that they are applicable to many different disciplinary contexts, including computing education. We are, however, interested in \emph{computing-specific pedagogies}, i.e., methods which have been developed in computing education contexts and can be applied (almost) only there.  Examples of such methods include pair programming~\cite{hanks2011pair,hawlitschek2023empirical}, Parsons problems~\cite{denny2008evaluating,du2020review}, computer science unplugged~\cite{battal2021computer,CSunplugged} or Use-Modify-Create ~\cite{lytle2019use}. We are not aware of any comprehensive list of computing-specific pedagogies. \citet{falkner2019Pedagogy} presented a deep discussion on a number of pedagogical approaches; however, they did not build a comprehensive list. \citet{sanders2017folk} discussed active learning approaches listing many specific pedagogies, supporting these in some way. However, the computing education education literature is vast, presenting a multitude of reported practices that are often tied to a certain context, depending on the teacher's experience, available teaching and learning resources (both hardware and software as well as physical environment), schedule and target student group.   We acknowledge that building a comprehensive list of computing-specific pedagogies is likely not possible due to the scope of literature published in over 50 years.  However, it is feasible to build a representative list which we set as our goal. We therefore proceeded, as follows.

\begin{enumerate}
    \item We collected pedagogies discussed in the previously mentioned papers~\cite{falkner2019Pedagogy,sanders2017folk}.
    \item We searched systematically through five years of papers (2018-2022) published in the ITICSE, SIGCSE, and ICER conferences as well as in ACM TOCE and CSE journals, to identify any new computing-specific pedagogical practices and techniques. We read through the title, abstract, and keywords to identify candidates, and then the full paper to describe them.  Each paper was analyzed by one author, who identified relevant candidates for such pedagogies. The candidates were discussed jointly to decide  whether they 
    could be applicable in a variety of computing education contexts. The criteria for ruling out a pedagogy were if the pedagogy required use of some specific software, which had been developed for local needs and was likely not accessible by others, or if the pedagogy was tailored for some special group of students, for example, based on their cultural knowledge and background.  We also excluded pure professional practices that are used in education, such as test-driven development that can be considered more as learning goals of professional skills than pedagogical practices.  We also excluded papers focusing on micro level pedagogies, such as a single or set of assignments.

    \item The summary list was augmented with expert knowledge of computing-specific pedagogies by three authors, all of whom had 10-30+ years of experience in teaching computing courses as well as an almost equally long experience in publishing in computing education research venues and who were thus highly knowledgeable of various pedagogical approaches that had been used and reported in literature.
    \item The final list covered 23 pedagogies, which are listed in Appendix A.  For each of these, we tried to find a paper where the pedagogy was first published.  However, this was not successful in all cases, and so we then identified an early paper that was widely cited in Google Scholar. We call these papers \emph{pedagogy source papers}.
\end{enumerate}

\subsection{Identifying computing-specific theories}

Our next goal was to identify a list of computing-specific theories in computing education. Here we built on the work by Malmi et al. who had explored these in several papers~\cite{malmi2019CompEdTheories,malmi2020TheoriesOfEmotion,malmi2022Developing} identifying well over 100 constructs in papers published in ICER, TOCE and CSE during 2005-2020, which they called \emph{theoretical constructs}. This list was augmented by the work by~\citet{szabo2023LearningTheories} who had also analyzed the use of learning theories in computing education. However, they used a broader search identifying additional computing-specific theoretical constructs.  When reviewing this list, we observed that many of the identified constructs reported in \cite{malmi2019CompEdTheories,malmi2020TheoriesOfEmotion,malmi2022Developing} were early developments, such as statistical models, grounded theories or phenomenographical outcome spaces, while the list in~\cite{szabo2023LearningTheories} had theories with  more broadly recognized names, such as \emph{engagement taxonomy}~\cite{Naps2002ExploringRole} or \emph{learning edge momentum}~\cite{robins2010learning}. Considering our main goal of finding connections between computing education specific theories and pedagogies, we decided to focus on more established theories or models for which we could identify a name,  either proposed by the original authors or used in other citing papers later on.  

In order to test whether this approach seemed promising, we used these names as keywords, searching them from the Scopus data base for their occurrence in titles, author keywords or abstracts.  It seemed likely that if they were recognized with a name, we would find hits for them in the data base. Indeed, for most of them, we found many hits, while some had none. We therefore decided to focus on a list of 21 theories for which we found hits in Scopus, and use these in further analysis.  These theories and models are listed in Appendix B. For each of them, we identified the original paper where it was published, as they were reported in \cite{malmi2019CompEdTheories,malmi2020TheoriesOfEmotion,malmi2022Developing,szabo2023LearningTheories}. We refer to these papers as \emph{theory source papers}.
 


\subsection{Identifying connections of theories and pedagogies}

In order to identify connections between computing education specific theories and pedagogies, we used the lists of pedagogy and theories source papers, as follows.
First, we identified all papers in the ACM Digital Library that cite a theory source paper, and correspondingly a pedagogy source paper. Finally, we cross-tabulated these papers with a script to identify all papers that cite one or more papers in both lists.  We refer to these papers  as \emph{intersection papers}.

A paper may cite both a theory source paper and a pedagogy source paper for multiple reasons. We therefore developed a categorization scheme that  would describe various ways a theory and pedagogy could be discussed and possibly connected in the paper. The initial categorization was revised several times when we analyzed the intersection papers, finally leading to the following categories (Table~\ref{tab:categories}).

\begin{table}[htbp]
    \centering
    \caption{Theory-Pedagogy Interaction Categories}
    \begin{tabular}{p{3cm}p{10cm}c}
        \toprule
        Category & Description & Level \\
        \midrule
        Separate discussion & \begin{tabular}[t]{@{}p{10cm}@{}}Theory and pedagogy are casually referenced in the paper or list of references, or possibly discussed in related work but there is no connection to the empirical study in the paper.\end{tabular} & 1 \\
        Discussed in context without relationship & \begin{tabular}[t]{@{}p{10cm}@{}}Theory and pedagogy are both discussed in relation to the empirical work reported in the paper, but no relationship between them is mentioned.\end{tabular} & 2 \\
        Explicitly connected in context & \begin{tabular}[t]{@{}p{10cm}@{}}The theory and pedagogy are explicitly connected in the context of the reported empirical work.\end{tabular} & 3 \\
        Analysis & \begin{tabular}[t]{@{}p{10cm}@{}}Theory and pedagogy are both used in the analysis or discussion of results. Theory is used to explain results of a pedagogy.\end{tabular} & 4 \\
        Theory development & \begin{tabular}[t]{@{}p{10cm}@{}}Pedagogy is used to develop, support, or discount an existing theory.\end{tabular} & 5 \\
        Pedagogy development & \begin{tabular}[t]{@{}p{10cm}@{}}Theory is used to develop, support, or discount an existing pedagogy.\end{tabular} & 6 \\
        Artefact development & \begin{tabular}[t]{@{}p{10cm}@{}}Theory and pedagogy are both used to design or develop another new theory/model/framework/instrument/pedagogy.\end{tabular} & 7 \\
        \bottomrule
    \end{tabular}
    \label{tab:categories}
\end{table}

\emph{Note for categories 1-4.} A vast majority of the analyzed papers presented empirical work, and we investigated whether a theory or pedagogy had some relation to it.  However, a few papers were either review papers or papers focusing on theoretical development/discussion.  For these cases, we investigated whether the theory or pedagogy was explicitly discussed and possibly used in argumentation in the review or theoretical discussion.  For brevity, we do not separately discuss these cases from empirical papers when reporting the results.

To analyze the interaction papers, all four authors read each paper in the set of papers from a jointly agreed theory-pedagogy combination, classifying each paper according to one of the categories in the  Theory-Pedagogy Interaction Categories scheme.  Thereafter, each paper was jointly discussed by all authors until a consensus of the categorization was agreed upon.  In cases where the categorization scheme was refined, the previously analyzed papers were revisited to match the revised categories.

After this analysis, as a further verification, we sorted the analyzed papers based on their categorizations.  Thereafter all papers in categories 3-7 were revisited by all four authors to confirm that the papers in each category were aligned with the category definition. In this process, a few papers were re-categorized and discussed until a final joint agreement was reached.
       

\section{Results}
\label{sec:results}
We present our results below as relevant to each research question.
\subsection{RQ1: What computing-specific pedagogies are defined in CER literature?}
Our data collection of computing-specific pedagogies from a systematic search of five years of publications (2018-2022) in five venues (ITICSE, SIGCSE, ICER, TOCE and CSE), a search of two key publications~\cite{falkner2019Pedagogy,sanders2017folk}, and expert knowledge found over 100 pedagogical models and techniques. This list was refined to remove pedagogies that were deemed too specific, were reliant on a particular technology or were considered just training professional practices. The final list of 23 pedagogical models and techniques is presented in Appendix A. 

Most of the pedagogies we found are focused on models and techniques for teaching (20) with just three focused on assessment (\emph{executable exams}, \emph{in-flow peer review}, \emph{peer code review}). The pedagogies range from broad models of how a computing course would be taught (e.g., \emph{bootcamp} and \emph{hackathon}) or an underlying theme or emphasis in a course (e.g., \emph{socially-responsible computing} and \emph{computing for the social good}), to specific techniques used to teach a particular topic or skill (e.g., \emph{pair programming} and \emph{plans and goals}). 

The most common pedagogies we found are concerned with learning or assessment of programming (15). Two are focused on teaching fundamental computing concepts (\emph{computer science unplugged} and \emph{notional machine}) with two others focused on more advanced topics (\emph{scrumage} and \emph{structured mentorship}). A couple of recently developed pedagogies are specific to cybersecurity (\emph{adversarial mindset in cybersecurity} and \emph{cryptographic playground}).

\subsection{RQ2: What computing-specific learning theories and models identified by a name are discussed in CER literature?}
We formed a list of computing domain-specific theories from Malmi et al.~\cite{malmi2019CompEdTheories,malmi2020TheoriesOfEmotion,malmi2022Developing} and ~\citet{szabo2023LearningTheories}. The list was refined to include only theories that are identifiable by name. The final list of 21 theories is presented in Appendix B. 

Using a classification of theory area of focus developed by  \citet{malmi2019CompEdTheories} we found that the theories covered 8 of the 11 focus areas defined by Malmi et al. The most common focus area was \emph{learning and understanding} (8), with three others closely connected with learning (\emph{learning behaviour}(4),  \emph{performance/progression/retention} (2), and \emph{emotion/beliefs/attitudes/self-efficacy} (2)). The remaining five are spread over the areas of \emph{computing education research} (2), \emph{contents/curriculum/learning goals} (1), \emph{errors/misconceptions} (1), and \emph{teaching/pedagogical content knowledge} (1). A surprising finding was there were none in the area of \emph{assessment/self-assessment}.

\subsection{RQ3: What kind of connections have been made between domain-specific theories and pedagogies in computing education?}
The analysis of intersections of 21 CE theories and 23 CE pedagogies produced 405 intersections. There were five theories that did not intersect with a pedagogy and seven pedagogies that did not intersect with a theory. These are indicated with an asterisk in the lists in Appendixes A and B. 

A table showing the frequencies of interactions of the 16 intersecting theories and 16 intersecting pedagogies is shown in Table \ref{table:intersections} (Appendix C). 
We did not include a further 234 cases where \emph{notional machine} as a theory  intersected with \emph{notional machine} as a pedagogy. This intersection is shown as an 'x' in Table \ref{table:intersections}. An investigation of a sample of these cases indicated that there were many papers where the notional machine had been reported as just a theoretical construct or a pedagogy and there was no intersection of theory and pedagogy. We left this analysis for future work.

The most common intersecting theories, producing 60\% of all intersections, were notional machine (97), learning edge momentum (76) and theory of instruction for introductory programming skills (70).

The two pedagogies that intersected most with theories were pair programming, which intersected 97 times over 12 different theories and Parsons problems, which intersected 61 times with over 11 different theories. Pair programming and Parsons problems produced 39\% of all the intersections. We focused on pair programming and Parsons problems for more detailed analysis.

\subsection{RQ3a What type of connections have been made between domain-specific theories and pair programming?}

\begin{figure*}[htbp]
  \centering
      \includegraphics[scale=0.5]{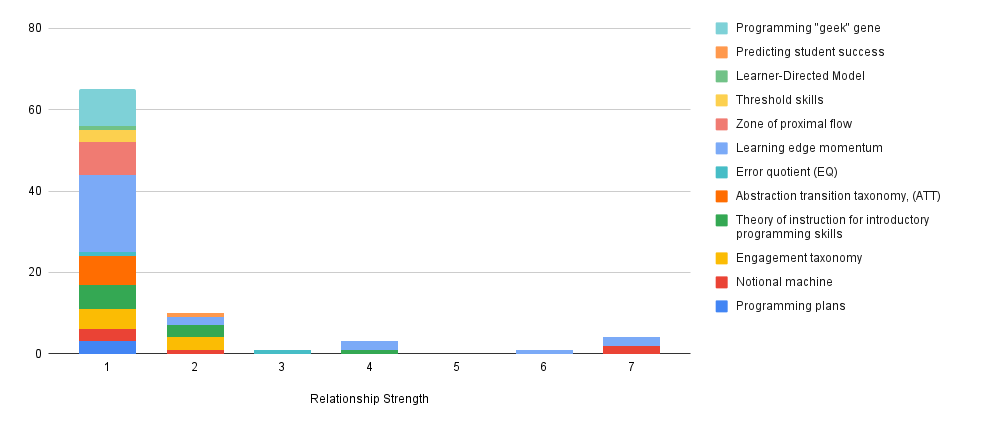}
\vspace{-12pt}  \caption{Connection types between computing-specific theories and Pair programming}
  \label{fig:pair} 
\end{figure*}

We classified each paper in the set of intersections of pair programming with a theory according to the theory-pedagogy interaction level as described in Table \ref{tab:categories}. From the 97 papers in the data set,  two papers were eliminated as they were duplicates or did not include a reference to the theory or pedagogy in the main text, with the remaining 95 describing intersections between pair programming and theories. A large majority of the intersections fell into categories 1 and 2 as shown in Figure~\ref{fig:pair}, denoting that both the pedagogy and the theory were only casually mentioned in the paper, or discussed separately with no connection to the empirical work in the paper. The frequencies of each category of interaction is shown in Table \ref{table:intersectionsPairProgramming}.

We found that pair programming intersected with 12  theories; however, only five theories had intersections at a level of 3 or above. There were no intersections with cognitive complexity of computer programs, normalized program state model (NPSM), progression of early computational thinking model (PECT) or reducing abstraction. We will now describe the cases where we found the strong connections between the pair programming and theory. 

\begin{table*}[h]
\centering
\caption{Counts of interactions of Pair Programming and theories at different levels of interaction: 3 Explictly connected in context; 4 Analysis; 5 Theory development; 6 Pedagogy development; 7 Artefact development.}
\label{table:intersectionsPairProgramming}
\begin{tabular}{l r r r r r r r}
\hline
 & \multicolumn{7}{c}{\textbf{Categories of Interaction}}\\
\textbf{Theory / model name} & \textbf{<3} & \textbf{3} & \textbf{4} & \textbf{5} & \textbf{6} & \textbf{7} & \textbf{Total}\\ \midrule
Abstraction transition taxonomy (ATT) & 8 & - & - & - & - & 1 & \textbf{9}   \\
Engagement taxonomy & 8 & - & - & - & - & - & \textbf{8}   \\
Error quotient (EQ)   & 1 & 1 & - & - & - & - & \textbf{2}  \\
Learner-Directed Model   & 1 & - & - & - & - & - & \textbf{1}  \\
Learning edge momentum   & 21 & - & 2 & - & 1 & 2 & \textbf{26}  \\
Notional machine   & 12 & - & - & - & - & 2 & \textbf{14}  \\
Predict student success (PreSS\#)  & 1 & - & - & - & - & - & \textbf{1}  \\
Programming "geek" gene   & 9 & - & - & - & - & - & \textbf{9}  \\
Programming plans   & 4 & - & - & - & - & - & \textbf{4}  \\
Theory of instruction for introductory 
programming skills   & 9 & - & 1 & - & - & - & \textbf{10}  \\
Threshold skills   & 3 & - & - & - & - & - & \textbf{3}  \\
Zone of proximal flow   & 8 & - & - & - & - & - & \textbf{8}  \\
\textbf{Total}    & \textbf{85} & \textbf{1} & \textbf{3} & \textbf{-} & \textbf{1} & \textbf{5} & \textbf{95}  \\
\hline    
\end{tabular}
\end{table*}

\subsubsection{Artefact development (Category 7)}
Pair programming has been strongly connected with several computing-specific  theories, namely, abstraction transition taxonomy (ATT), learning edge momentum (LEM), and notional machines (NM), resulting in various artefact developments. ATT represents a foundation theory for PRIMM \cite{sentance2019teaching}, where three key principles are proposed to guide the teaching of programming, namely, mediation through language, the understanding that learning moves from the social plane to the cognitive plane, and the role of the \emph{more knowleadgeble other} (MKO) in the Zone of Proximal Development \cite{basawapatna2013zones}. Mediation through language suggests that students should be encouraged to work together and discuss through a social construction of knowledge.  Pair programming is a pedagogical technique suggested by the authors as an implementation of this first principle. In \cite{chetty2015towards}, pair programming and LEM are used in the development of the pedagogical design for CS1 in a South African university, providing avenues for struggling novice programmers to improve their skills through pair programming. Specifically, as guided by LEM, the proposed approach focused on identifying quickly whether learning was successful, and individual one-on-one support was provided to students who did not succeed in the first weekly assignment. Pair programming was one of the main pedagogies used in the course, together with teaching code reading before writing, and teaching by focusing on time-on-task. Deconstructionism is proposed as a pedagogy in \cite{griffin2016learning}, where learners spend as much time deconstructing code as they do writing code. Notional machines form a backbone for this new pedagogy, and pair programming activities are a type of activity that fits naturally within the proposed pedagogy, among others suggested by the author, such as peer instruction and POGIL.

\subsubsection{Pedagogy development (Category 6)}
Wood et al. \cite{wood2013s} use LEM to develop best practices for pair programming, specifically by framing the concept of programming confidence, which is defined as the way in which students approach programming exercises and distinguishes between confident students who are willing to experiment with techniques and are unfazed by coding errors, and students who are unable to make independent progress and frequently become 'stuck'. Using LEM and the concept of confidence, authors propose to group students by confidence in their pair programming exercises, and to monitor those students who do not gain confidence very early in the course, in order to provide support.

\subsubsection{Analysis (Category 4)}
\textit{Analysis} relationships were identified between pair programming and LEM and Xie's theory of instruction for introductory programming skills \cite{xie2019theory}. Both pair programming and LEM are used in the theory-based discussion of feedback in \cite{ott2016translating}. Hausswolff \cite{von2022practical} introduces several theoretical underpinnings based on Dewey \cite{dewey1974john}, Wittgenstein \cite{williams1999wittgenstein} and Deleuze \cite{rodowick1997gilles} for learning how to program, which are then considered in the analysis of results and experiences of students doing pair programming exercises. The discussion considers both the pair programming undertaken by students under a LEM lens, in particular when attempting to understand the challenges faced by novice programmers. \citet{druga2022families} analyse how families designed and developed program games during an in-home study. Xie's theory of instruction for introductory program skills \cite{xie2019theory} is used to design the analysis codes, and pair programming is recognised as one of the behaviors undertaken by the families building the games, and its dynamic is used to analyse the study results.  

\subsubsection{Explicitly connected in context (Category 3)}
Lastly, pair programming and learning edge momentum and error quotient theory are explicitly connected in context, respectively, in \cite{joshi2018reflecting}, focused on training teachers in inclusive practice, and in \cite{rodrigo2013analysis}, focused on understanding programming behaviors of differently skilled programmers.

\subsection{RQ3b: What type of connections have been made between domain-specific theories and Parsons problems?}

\begin{figure*}[htbp]
  \centering
      \includegraphics[scale=0.5]{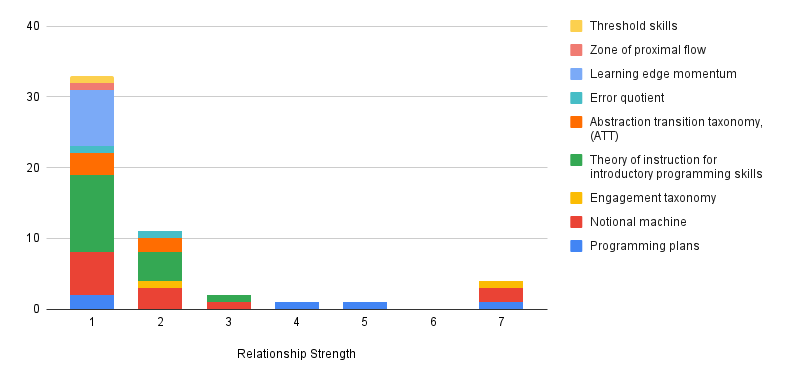}
  \vspace{-10pt}
  \caption{Connection types between computing-specific theories and Parsons problems}
  \label{fig:Parsons} 
  
\end{figure*}

We classified each paper in the set of intersections of Parsons problems with a theory according to the theory-pedagogy interaction level as described in Table \ref{tab:categories}. From the 61 papers in the data set, four papers were eliminated as they were duplicates or did not include a reference to the theory or pedagogy in the main text, with the remaining 57 describing intersections between Parsons problems and theories. Similarly to pair programming a large majority of the intersections fell into categories 1 and 2 as shown in Figure~\ref{fig:Parsons}. The frequencies of each category of interaction is shown in Table \ref{table:intersectionsParsons}.

We found that Parsons problems intersected with 11 theories; however, only four theories had intersections at a level of 3 or above. There were no intersections with cognitive complexity of computer programs, learner-directed model, normalized program state model (NPSM), predicting student success, progression of early computational thinking (PECT) model or reducing abstraction. We will now describe the cases where we found the strong connections between Parsons problems and theory.

\begin{table}[h]
\centering
\caption{Counts of interactions of Parsons Problems and theories at different categories of interaction: 3 Explictly connected in context; 4 Analysis; 5 Theory development; 6 Pedagogy development; 7 Artefact development }
\label{table:intersectionsParsons}
\begin{tabular}{l r r r r r r r}
\hline
 & \multicolumn{7}{c}{\textbf{Categories of Interaction}}\\
\textbf{Theory / model name} & \textbf{<3} & \textbf{3} & \textbf{4} & \textbf{5} & \textbf{6} & \textbf{7} & \textbf{Total}\\ \midrule
Abstraction transition taxonomy (ATT) & 5 & - & - & - & - & - & \textbf{5}   \\
Engagement taxonomy  & 1 & - & - & - & - & 1 & \textbf{2}   \\
Error quotient (EQ)   & 3 & - & - & - & - & - & \textbf{3}  \\
Learning edge momentum   & 8 & - & - & - & - & - & \textbf{8}  \\
Notional machine   & 9 & 1 & - & - & - & 2 & \textbf{12}  \\
Programming "geek" gene   & 1 & - & - & - & - & - & \textbf{1}  \\
Programming plans   & 2 & - & 1 & 1 & - & 1 & \textbf{5}  \\
Theory of instruction for introductory programming skills   & 16 & 1 & - & - & - & - & \textbf{17}  \\
Threshold skills   & 1 & - & - & - & - & - & \textbf{1}  \\
Zone of proximal flow  & 3 & - & - & - & - & - & \textbf{3}  \\
\textbf{Total}  & \textbf{49} & \textbf{2} & \textbf{1} & \textbf{1} & \textbf{-} & \textbf{4} & \textbf{57}  \\
\hline    
\end{tabular}
\end{table}

\subsubsection{Artefact development (Category 7)}
The pedagogy of Parsons problems  has been strongly connected within an \emph{artefact development} relationship with several theories, namely, programming plans \cite{cunningham2021avoiding}, notional machine \cite{izu2019fostering, clements2022towards} and engagement taxonomy \cite{al2019social}. Cunningham et al.  \cite{cunningham2021avoiding}  use programming plans and Parsons problems to design a purpose-first programming approach to teach conversational programmers, focusing on learning a handful of domain-specific code patterns and assembling them to create authentic programs. They develop a purpose-first programming prototype that teaches five web scraping patterns. Purpose-first programming enforces a sequence between subgoals driven by variables, where a subgoal uses previously defined  variables and/or produces variables to be used in later subgoals. In the development of the artefact, the authors are inspired by Parsons problems devise a three-part activity, in which (i) learners pick from a bank of plan of goals and arrange them in the correct order, (ii) learners repeat the previous activity but with plan code, and (iii) learners fill in the the slots in the code they have assembled. In \cite{izu2019fostering}, notional machine and Parsons problems are both used to create new program comprehension activities and to propose new learning trajectories. In \cite{clements2022towards}, a configuration of Snap! is developed using notional machine and Parson problems.  Similarly, Parsons problems and engagement taxonomy are used in \cite{al2019social} to create social worked examples, in a reciprocal connection. Specifically, engagement taxonomy provides guidelines for high engagement, which in  turn enhances the effectiveness of Parsons problems.  

\subsubsection{Theory development (Category 5)}
A \textit{theory development} relationship was recorded between programming plans and Parsons problems \cite{malysheva2020using}. The authors  identify "plan structure errors" when analyzing students' solutions to  Parsons problems that are used as a pedagogical activity. Aided by the programming plan error taxonomy \cite{cunningham2021avoiding}, the authors define plan structure errors based on what is wrong with the specific placement of the line in the Parsons problem, as \textit{plan structure errors} when pair of actions animate out of order, \textit{control flow error} when the line is placed inside the wrong parent control flow block, \textit{do both}, when an interaction error occurs, \textit{do nothing}, when the error involves an empty control flow structure that is misplaced relative to another line.

\subsubsection{Analysis (Category 4)}
An analysis relationship was found between Parsons problems and notional machine when analysing unplugged activities \cite{munasinghe2023unplugged}. Notional machines form a backbone of the unplugged activities under analysis, and programming plans were used to analyse whether the unplugged activities differ from their plugged counterparts, in both computing and mathematics. The use of unplugged activities exposes learners to an accessible and explicit notional machine.  Notional machine continues to provide support for programming tasks, that are introduced through a careful progression of activities.

\subsubsection{Explicitly connected in context (Category 3)}
Lastly, Parsons problems and the theory of instruction for introductory programming skills \cite{xie2019theory} were explicitly connected in context in \cite{fowler2022reevaluating}, as Parsons problems is a specific exercise type that can be used to explain the relationship between explaining, tracing, and writing skills that are critical within Xie's theory.

\section{Discussion}

We discuss first our experience in identifying pedagogies and theories and then comment on the results.

\subsection{Identifying pedagogies, theories and models}

We set our goal to identify a list of computing-specific pedagogies. The task turned out be complex for several reasons.  Firstly, pedagogical practices are flexible concepts. They need to be adaptable to different teaching and learning contexts, depending on the specific learning goals, target population as well as teaching and learning resources. In this sense, they cannot too rigidly define the interplay of the relevant factors: activities, students' and teacher(s)' roles and requested resources. Rather, they need to include principles and guidelines how these factors should be used, and teachers will organize the actual implementation of the pedagogy in a particular setting. Therefore, the level of descriptions in publications, how some practice was implemented varied significantly.  

Secondly, some pedagogies that have emerged in computing education could also be adapted to other disciplines, such as Structured mentorship, Socially-responsible computing, Computing for social good and In-flow peer review. It is actually plausible that similar type of high-level pedagogies have been developed and used in other disciplines, too. Thus, the borderline between computing-specific and generic pedagogies is not strict. Thirdly, an inherent part of computing education is training some professional practices. For example, if programming courses involve project work where designing programs using UML notation or implementing software using test-driven development is requested, are these pedagogical methods or simply activities to learn a specific professional skill?  We deemed the activity itself not a pedagogy in the same way as essay assignment writing is not a pedagogy but just a professional practice. However, training a skill using pedagogically tailored tasks and actions could be considered a corresponding pedagogy, but such details are often missing from published papers. 
Fourthly, we deliberately ruled out pedagogical practices which require certain specific software to be used. Computing educators have developed numerous software tools to support learning, e.g.,\cite{luxton2018introductory}.  However, these tools are very often tailored to match a local context and they may be hard to access.  Moreover, they often have little if any support or maintenance for external 
users~\cite{malmi2019tools} making pedagogies built on them vulnerable.  


General pedagogies can be applied in multiple disciplinary contexts, while context-specific pedagogies have some components which are (almost) unique  for a certain discipline.  In computing education, programming is a unique core activity for learning the disciplinary skills.\footnote{Programming tasks are, of course, used in many other disciplines, too.  However, learning programming is part of computing discipline.} Moreover, programming is a relevant activity in implementing a large share of advanced computing topics and skills, and it covers multiple computing-specific skills, such as program design, testing, debugging, performance analysis, software architecture design etc. It is thus understandable that most of the computing-specific pedagogies we found are related to teaching programming.

The search for computing-specific theories was more straigthforward, as we could build on existing theory/model lists from literature. 
Most of the computing-specific theories and models that we deemed more established ones (having a name or an acronym) were related to programming, too.  

\subsection{Combining theories and pedagogies}

As CER is developing and growing as a field of science, it is natural that there is a growing trend of building theories and models which are based on the discipline itself~\cite{malmi2022Developing}. Thus, we expected to find some interplay of the theories and pedagogical practices.  This could happen in many ways. \citet{malmi2023domain_practice} had explored this interplay from the perspective that theoretical constructs had been used to develop any new pedagogical practices, finding a small number of cases either in papers citing the theory source paper or within the discussion of the source paper itself.  They recommended that research papers should more frequently have such discussion of "pedagogical implications" which could make the results more accessible to broader computing educator audience and not only for active computing education researchers. Compared to their goals, our perspective was to explore the interplay of existing pedagogies and theories, which could take place in multiple forms.  We indeed found many, over 400 interactions. A detailed analysis of all of them was beyond the scope of this paper and we therefore focused on the most common cases which were related to Pair programming and Parsons problems which we investigated more closely. Other pedagogies with rich interaction with theories were Notional machine (as a pedagogy), Program tracing, Computer science uplugged, Live coding, Media-based computation and Peer code review, all of which are related to learning programming. The most commonly intersected theories were Notional machine (as a theory), Theory of instruction for introductory programming skills, Learning edge momentum, all related to programming, as well.

The closer analysis of pair programming and Parsons problems revealed that a great majority of the intersections were on levels 1 and 2 implying that there was no clear connection between the theory and pedagogy in the paper.  This is no surprise, as papers are cited for many reasons, most often simply as related work. The more interesting cases were on levels 3-7, especially on the upper levels.  PRIMM pedagogy~\cite{sentance2019teaching} is a good example. It is explicitly building on the theoretical understanding of how teachers and students discuss in three different levels of language (English, CS Speak, and Code), as the Abstraction transition taxonomy~\cite{cutts2012abstraction} presents. To encourage discussion on multiple levels, which support students to grow into the computing community, the pedagogy integrates heavily pair programming as an activity where students naturally need to discuss much. \citet{clements2022towards} on the other hand, seeking to support students to understand notional machine and runtime stacks, were inspired by Parsons problems as a method. They planned a similar tool, implemented with Snap!, where students can work with a interactive visual presentation of stack frames and better understand how they work.  These kinds of explicit combinations, where theory provides a suggestion what kind of pedagogical practices would be effective or a pedagogy inspires a new way of learning theory are something which we hope to see more.  

Theory and pedagogy can support each other in different ways.  Pedagogy can have an impact which matches a theory, thus providing some evidence to support the theory.  On the other hand, theory can suggest ways to improve the implementation of an existing pedagogy.  Learning edge momentum theory\cite{robins2010learning} suggests that due to the scaffolded structure of programming knowledge, students who face problems in the beginning of CS1 may increasingly fall behind the others and are in risk of dropping out. Thus, supporting  especially weaker students is important. Using best practice from pair programming research, \citet{wood2013s} proposed using programming confidence (interpreted as student's current programming skill) as an indicator how to pair students (with similar confidence) as well as switching pairs regularly, finding out that students with low confidence benefited from this practice and increased their confidence.

While our closer analysis focused only on pair programming and Parsons problems, we found many interesting examples of how computing-specific theories/models and pedagogies were used together to further research and improve education. We are confident that analysing the remaining intersection cases would reveal more such cases. This encourages to seek research designs where the pedagogy is explicitly selected or tuned match the theory, or theory is explicitly used to inform pedagogical choices.


\subsection{Limitations}
\label{subsec:limitations}

Our search for candidate pedagogies covered two key references~\cite{falkner2019Pedagogy,sanders2017folk} and five recent years of publications in major computing education research venues, complemented with expert knowledge from three authors with a long experience in computing education and CER. For each of the candidates in the search we discussed it with at least three authors to find consensus whether it could be accepted based on the above criteria. We acknowledge that even in this way, the inclusion/exclusion decision was not clear cut.  Therefore we considered the consensus method useful.

We acknowledge that broader search of years and venues would have increased the number of identified pedagogical practices. However, this was beyond the scope of our resources, as it would cover thousands of papers. Moreover, not all well-functioning pedagogies are reported in published papers, and talented teachers have certainly invented many practices which work well in their contexts.  However, there is no obvious way to collect such pedagogical content knowledge in a large scale. Surveys to mailing lists and interviews with colleagues would produce some results, but likely many findings would anyway overlap with what we identified in our literature search.

Despite these challenges, we consider building a list of computing-specific pedagogies a meaningful effort, which can be helpful to many readers as a source of educational resources. We acknowledge that such a list is never complete, as the field evolves constantly. For example, the fast developing research related how large language models can be used to support education will certainly create new pedagogical methods, such as prompt problems to guide AI tools to build programs~\cite{denny2024prompt}. 

  The decision to focus on theories with some name or acronym was supported by our trial search in Scopus.  We considered it plausible that if a theory/model had some clear role in a paper, it would likely be mentioned in the title, abstract or author keywords.  We acknowledge that this does not apply to all papers which cite a theory source paper.  However, \citet{malmi2022Developing} found that a vast majority of papers citing a theory source paper does not use the construct; the paper is likely cited for some other reason. We considered that based on this finding, we likely would not miss many cases relevant to our goals.

The intersection search for pedagogies and theories was carried out in ACM digital library only, which --- to our conception --- covers significant share of international computing education research.  Searches for other data bases were therefore excluded to reduce the total workload.  Finally, the categorization of theory-pedagogy intersections were carefully carried out by all 4 authors until a consensus of the category for each paper was reached.  Thereafter we carried out a second pass comparing papers within a single category with each other until the final consensus for all papers on levels 3-7 were reached.  We left out closer analysis of levels 1 and 2 papers, because they were considered uninteresting from the purpose of this work.

\section{Conclusion}

Computing education has developed its own pedagogical practices for decades. Theoretical development is considerably more recent activity and there is growing interest in it~\cite{malmi2022Developing}. We have explored the interplay of these field-specific pedagogies and theories/models in CER literature identifying many interesting examples where they support each other in different ways.  We therefore recommend further work to combine them. Field-specific theories and models can provide valuable arguments for developing new pedagogical practices or further development of existing ones.  Instead of building on implicit assumptions, they can provide arguments which are based on empirical results and thus support systematic development of computing education. 


In this paper, we have analyzed only intersections of computing-specific pedagogies and theories, as we believe that their combinations are essential to further computing education research and practice. We continue the detailed analysis of theory-pedagogy intersections for other pedagogical practices than pair programming and Parsons problems, which were now left out from this paper. However, there is also much space in future work to explore similar connections between computing-specific and generic pedagogies and learning theories. We hope to see rich work in these areas with the future view where computing educators would have more explicit arguments why they select some pedagogical choices for their contexts. This might sometimes cause questioning their current implicit assumptions, but if not, making assumptions visible and explicit allows better opportunities to understand what works and why in pedagogy. 

Finally, we concur with~\citet{malmi2023domain_practice} and recommend CER publication venues to encourage authors to discuss pedagogical implications as a regular part of discussion, where this is relevant.  This would help practicing educators to learn from the research and thus disseminate relevant findings for wider audience.



\begin{table*}[hbpt]
\centering
\begin{tabular} {p{0.04\linewidth} p{0.25\linewidth} p{0.63\linewidth} p{0.04\linewidth} r}

\multicolumn{4}{p{1.0\linewidth}}%
{\textbf{Appendix A} Computing education pedagogical models or techniques identified in the study; ordered alphabetically. Asterisks denote that we found no intersection between the pedagogy and some theory in Appendix B.} \\

\toprule  \textbf{Id} & \textbf{Pedagogical model or technique}  & \textbf{Explanation} & \textbf{Paper}\\ \midrule 
P1* & Adversarial mindset in cybersecurity & A cybersecurity course design that balances theoretical and practical learning with emphasis on exploring offensive tactics, techniques, and procedures & \cite{oconnor2022helo}\\
P2 & Bootcamp & A intensive, specialized, short term training course designed to rapidly prepare students for entering the software industry  & \cite{thayer2017barriers}\\
P3 & Computer science unplugged & An approach to teaching computer science concepts through games and puzzles rather than on a computer & \cite{bell2009computer}\\
P4 & Computing for social-good & A teaching approach using educational activities that emphasize computing for the social good  & \cite{goldweber2019computing}\\
P5* & Cryptographic playground & An online environment where students can experiment with and learn about cryptosystems & \cite{lodi2022cryptography}\\
P6* & Executable exams & An exam conducted on computer in a programming environment where student can write, compile, run and test their programs & \cite{bourke2023executable}\\
P7 & Hackathon & A fast-paced event where people work collaboratively in teams  to create a software application over short period of time & \cite{nandi2016hackathons}\\
P8 & In-flow peer review (IFPR) & A peer-review conducted during the development of an assignment  & \cite{clarke2014flow}\\
P9 & Live coding & A teaching technique where code is written and tested live on a computer in front of students during a class & \cite{selvaraj2021live}\\
P10 & Media-based computation & An approach to teaching introductory computing with a media-focused context & \cite{guzdial2013exploring}\\
P11 & Notional machine & A teaching approach that uses the explicit concept of a notional machine to explain programming constructs or semantics & \cite{dickson2020engage}\\
P12 & Pair programming & A teaching technique where two students work side-by-side at one computer, continuously collaborating on the development of the same program  & \cite{beck2000extreme}\\
P13* & Pair-separate, pair-together, and partner puzzles & A categorisation of three different collaboration modes of novice programmers in a block-based environment  & \cite{lytle2019use}\\
P14 & Parsons problems & An instructional tool for introductory programming where students piece together programming solutions from code fragments  & \cite{Parsons2006parson}\\
P15 & Peer code review  & A collaborative learning approach where students peer review programs written by other students and give feedback & \cite{wang2012assessment} \\
P16 & Plans and goals & An approach to teaching programming that incorporates explicit use of plans and goals & \cite{de2009teaching}\\
P17* & Pre-programming Analysis Guided Programming (PAGP)  & A structured process to guide students in mimicing how an expert approaches a programming task  & \cite{jin2008pre}\\
P18 & Program tracing & A technique where the programmer traces the execution of program code in order to track the values of variables during execution and to determine the output of the code  & \cite{lopez2008relationships}\\
P19 & Scrumage (SCRUM for AGile Education) & An agile teaching approach that aims to mimic workplace expectation and promote student autonomy  &\cite{duvall2021improving}   \\
P20* & Socially-responsible computing; green education & A teaching model that exposes students to the social impact and ethics of computing & \cite{cohen2021new}\\
P21* & Structured mentorship & A mentorship program that combines industry expertise, peer mentorship, and relevant skills, to complement classroom learning  & \cite{mbogo2019structured}\\
P22 & Tangible; kinesthetic; physical computing & An approach to teaching programming where students arrange and connect physical objects to represent various programming elements, forming physical constructions that describe computer programs  & \cite{horn2019tangible}\\
P23 & Use-modify-create (UMC) & A scaffolding framework describing three phases of a learning progression where students use other's code, modify code and then create code & \cite{lee2011computational} 

\end{tabular}
\end{table*}

\begin{table*}[hbpt]
\centering
\begin{tabular}{p{0.25\linewidth} p{0.71\linewidth} p{0.04\linewidth} r}

\multicolumn{3}{p{1.0\linewidth}}%
{\textbf{Appendix B} Computing education domain-specific theories and models identified in the study; ordered alphabetically. Asterisks denote that we found no intersection between the theory and some pedagogy in Appendix A.} \\

\toprule  \textbf{Theory / Model name}  & \textbf{Explanation} &\textbf{Paper} \\ \midrule 

*2DET engagement taxonomy & Extending the engagement taxonomy for program/algorithm visualization to incorporate content ownership dimension & \cite{sorva2013review} \\

Abstraction transition taxonomy, ATT & Classification scheme for the knowledge and practices required to apprentice students into the programming community & \cite{cutts2012abstraction} \\

Cognitive complexity of computer programs (CCCP) & Framework characterizing the complexity of a program from a cognitive perspective in terms of the hierarchical structure of plans present and their interactions &  \cite{duran2018towards} \\

*Didactic Focus-based Categorization Method (DFCM)  & Scheme for classifying literature based on didactic focuses of educational research & \cite{Kinnunen2010HaveWeMissed} \\
 
Engagement taxonomy & Taxonomy of students' interaction with algorithm visualization  &  \cite{Naps2002ExploringRole}\\

Error quotient (EQ) & Measure of how much a student struggles with syntax errors while programming & \cite{jadud2006methods} \\

Learner-Directed Model & Pedagogical model combining Kolb's experiental learning and self-regulated learning & \cite{lee2016effectiveness} \\

Learning edge momentum & Theory to explain learning progression in introductory programming & \cite{robins2010learning} \\

Normalized program state model (NPSM) & Characterization of students’ programming activities in terms of the dynamically changing syntactic and semantic correctness of their programs &  \cite{carter2015normalized} \\

Notional machine & Idealized model of the computer implied by the constructs of the programming language & \cite{du1981black} \\

Predict student success (PreSS\#) & Machine learning model that can predict student success early in an introductory programming module (extension of PreSS) & \cite{quille2019cs1} \\

Programming "geek" gene & Hypothesis of an innate talent for programming & \cite{ahadi2013geek} \\

Programming plans & Expert knowledge as program fragments that represent stereotypic action sequences in programming & \cite{soloway1984empirical} \\

Progression of early computational thinking (PECT) model & Framework for understanding and assessing computational thinking in primary school & \cite{seiter2013modeling} \\
 
Reducing abstraction & Mental mechanism of reducing abstraction helps students to cope successfully with problems presented to them & \cite{hazzan1999reducing} \\

*Simon’s scheme & Scheme for classifying CER literature based on Context, Theme, Scope and Nature & \cite{Simon2007ClassificationAustralasian} \\

Theory of instruction for introductory programming skills & Holistic theory for teaching programming based on identification of four distinct skills that novices learn incrementally &  \cite{xie2019theory}  \\

Threshold skills & Extension of threshold concepts to include threshold skills & \cite{sanders2012threshold} \\
 
*Twelve Emotions in Academia Model & Main emotions detected in educational contexts & \cite{ruiz2016supporting} \\

*Weighted learning gain (WLG) & Measurement of student learning gains for isomorphic questions in the context of peer instruction & \cite{porter2011peerInstruction} \\

Zone of proximal flow & Pedagogical design framework integrating Vygotsky’s zone of proximal development theory with Csikszentmihalyi’s ideas about flow & \cite{basawapatna2013zones} \\







\end{tabular}
\end{table*}



\begin{landscape}
\begin{table*}[h]
\centering
\caption{\textbf{Appendix C}. Frequencies of the intersections between theories and pedagogies. Note that theories and pedagogies that had no intersections have not been included. The 'X' is used to indicate the intersection of notional machine as a theory with notional machine as a pedagogy.}
\label{table:intersections}
\begin{tabular}{c}
\hline
\textbf{Pedagogies}\\
\end{tabular}
\begin{tabular}{l r r r r r r r r r r r r r r r r r r r}
\textbf{Theory / model name} & \textbf{P2}  & \textbf{P3} & \textbf{P4} & \textbf{P7} & \textbf{P8} & \textbf{P9} & \textbf{P10} & \textbf{P11} & \textbf{P12} & \textbf{P14} & \textbf{P15} & \textbf{P16} & \textbf{P18} & \textbf{P19} & \textbf{P22} & \textbf{P23} & \textbf{Total}\\ 
\midrule
Abstraction transition taxonomy (ATT) & - &  2 & - & 1 & - & 2 & 2 & 6 & 9 & 5 &- & - & - &  - & - & 2 & \textbf{29}\\
Cognitive complexity of computer  \\
programs & - & - & - &  - & - & - & 1 & 1 & - & - & - & - & - & - & -  & - & \textbf{2} \\
Engagement taxonomy  & -  & 1 & - & - & - & - & 1 & 2 & 9 & 2 & 1 & - & 1 &  - & - & - & \textbf{17}\\
Error quotient (EQ)  & -  & - & 1 & 1 & - & 2 & 2 & 2 & 3 & 3 & 2 & - & 2 & -  & 2 & - & \textbf{20} \\
Learner-Directed Model  &  - & - & - & - & - & - & - & - & 1 & - & - & - &- &  - & - & - & \textbf{1}\\
Learning edge momentum  & -  & 3 & 1 & 1 & - & 4 & 6 & 13 & 26 & 8  & 9 & - & 4 & - & 1 & - & \textbf{76}\\
Normalized program state model (NPSM)  & - & - & - & - & - & - & - & - & - & - &  - & - & - & - & 1 & - & \textbf{1} \\
Notional machine  & 2  & 10 & 1 & 3 & - & 10 & 9 & x & 14 & 13 & 5 & - & 22 & - & 7 & 1 & \textbf{97} \\
Predicting student success  &  - & - & - & - & - & - & - & - & 1 & 1 &- & - & - & - & - & - & \textbf{2}\\
Programming "geek" gene  & -  & 1 & - & - & - & 1 & 2 & 2 & 9 & 1 & 2& - & - & - & - & - & \textbf{18}\\
Programming plans  & -  & - & - & - & - & 1 & - & 7 & 4 & 6 & 4 & 1 & 3 & - & - & - & \textbf{26}\\
Progression of early computational \\
thinking (PECT) model  & - & 1 & - & - & - & - & - & - & - & - & - & - & - & - & 1 & - & \textbf{2}\\
Reducing abstraction  & - & 3 & 1 & - & - & - & - & - & - & - & - & - & - & - & - & 1 & \textbf{5}\\
Theory of instruction for  \\
introductory programming skills  & 1  & - & - & - & 1 & 4 & 2 & 11 & 10 & 17 & 3& - & 19 & 1 & - & 1 & \textbf{70}\\
Threshold skills  & -  & - & - & - & - & - & 1 & 1 & 3 & 1 & 1 & - & 2 & - & - & - & \textbf{9} \\
Zone of proximal flow  & 2  & 5 & - & 1 & - & 2 & - & 3 & 8 & 4 & 2 & - & 1 & - & - & 2 & \textbf{30}\\
\textbf{Total} & \textbf{5}  & \textbf{26} & \textbf{4} & \textbf{7} & \textbf{1} & \textbf{26} & \textbf{26} & \textbf{48} & \textbf{97} & \textbf{61} & \textbf{29} & \textbf{1} & \textbf{54} & \textbf{1} & \textbf{12} & \textbf{7} & \textbf{405}\\
\hline    
\end{tabular}

\end{table*}
\end{landscape}

\bibliographystyle{ACM-Reference-Format}
\bibliography{CompPedagogiesTheory}
\end{document}